\documentclass{article}
\usepackage[left=1.25in,top=1.25in,right=1.25in,bottom=1.25in,head=1.25in]{geometry}
\usepackage{amsfonts,amsmath,amssymb,amsthm}
\usepackage{verbatim,float,url,enumerate}
\usepackage{graphicx,subfigure,epsfig,psfrag}
\usepackage{dsfont,bm,color,appendix}
\usepackage{natbib}

\usepackage{algorithm}
\usepackage{algpseudocode}

\def\one{{\bf 1}}

\usepackage[T1]{fontenc}

\title{A Pliable  Lasso}
\author{ Robert Tibshirani and Jerome Friedman \\
Departments of Biomedical Data Science and Statistics,\\ Stanford University }

\begin{document}
\maketitle

\begin{abstract}
We propose a  generalization of the lasso that allows the model coefficients to vary as a function of a general set of modifying
variables. These modifiers might be variables such as gender, age or time. The paradigm is quite general, with each lasso coefficient
modified by a sparse linear function of the  modifying variables $Z$. The model is estimated in a hierarchical fashion to control the
degrees of freedom and avoid overfitting. The modifying variables may be observed, observed only in the training set,
or unobserved overall. There are connections of our proposal  to varying coefficient models and high-dimensional interaction models.
We present a computationally efficient algorithm for its optimization, with exact screening rules to facilitate application to
large numbers of predictors.
The method  is illustrated on  a number of different simulated and real examples.  

\end{abstract}

\section{Introduction}

In this paper we consider the usual supervised learning problem.
  Given predictors $x_{ij}$ and response values $y_i$ for  $i=1,2,\ldots N$
and $j=1,2,\ldots p$, one popular method for this problem is the lasso.
This approach solves the  $\ell_1$-penalized  regression
 \begin{equation}
 {\rm minimize}\; \frac{1}{2}\sum_{i=1}^N(y_i- \beta_0-\sum_{j=1}^p x_{ij}\beta_j)^2+
 \lambda\sum_{j=1}^p|\beta_j|.
 \label{eqn:lasso}
 \end{equation}
 
This is equivalent to minimizing the sum of squares with constraint $\sum|\beta_j|\leq s$.
The absolute value penalty in (\ref{eqn:lasso}) induces sparsity in the solution for sufficiently
large values of $\lambda$; that is, some or many of the components of the solution $\hat\beta$ are zero.
The lasso and associated techniques like the elastic net (\citet{ZH2005}) are widely used. The R language package
{\tt glmnet} solves the lasso as stated  (\ref{eqn:lasso}) and in  a broader  class of  problems (such as generalized linear models)
very efficiently, using a coordinate descent procedure. Note that the objective function in (\ref{eqn:lasso}) is convex, making the optimization problem tractable.

In this paper we extend the lasso, embedding it in a more general model.
We suppose that in addition to our predictors and outcome, we have measurements of one or more ``modifying variables'' $z_{ik}$ for $i=1,2\ldots N $ and $k=1,2,\ldots K$.
For example, such a modifying variable might be sex (male or female), and we allow for  the possibility that some or  all of the coefficients $\beta_j$ are different for males and females.
Or $Z$ might be time and we wish to allow  time-varying coefficients.  Another possibility is to choose the modifying variables to be some  of the predictors $x_{ij}$.

Let $y$ be the $N-$vector of outcomes, and let $X, Z$  be $ N\times p$ and $N\times K$ matrices containing values of the predictor and modifying variables respectively. Let $X_j$ be the $j$th column of $X$ (an $N$-vector), and let $\bf 1$ be a column $N$-vector of ones.
The {\em pliable lasso} model has the form
\begin{eqnarray}
\hat y&=&\beta_0 \one+\Delta_0(Z)+  \sum_{j=1}^p X_j \circ (\beta_j \one+\Delta_j(Z)) \cr
&=&\beta_0 {\bf 1}+\Delta_0(Z)+  \sum_{j=1}^p (X_j \beta_j+ X_j \circ \Delta_j(Z))
\label{eqn:plasso}
\end{eqnarray}

In the above, $(\circ)$ is component-wise multiplication. In the first form, the $N$-vectors $\Delta_j(Z)$  are seen to modify the coefficients $\beta_j$;
the second form expresses  this equivalently as an interaction.
Note that if all of the $\Delta_j(Z)$ are zero, then (\ref{eqn:plasso}) reduces to the usual lasso.

For statistical power and interpretability, we also add an (asymmetric) weak hierarchy constraint:

\medskip

\begin{equation}
{\Delta_j(Z) \; {\mathbf {can \;be\; nonzero\; only\; if }\;} \beta_j \;{ \mathbf {is \;nonzero}}}.
\label{eqn:hier}
\end{equation}

\medskip

There are various possibilities for the form of $\Delta_j(Z)$.
One idea is to estimate each $\Delta_j(Z)$ using the predicted values from a regression tree fit to $Z$. This is attractive because regression  trees are simple and flexible, and can easily handle variables of different types. However the resulting
optimization problem is non-convex, making computations challenging. We discuss this approach briefly  in Section \ref{sec:further}.

In our main proposal we assume instead that $\Delta_j(Z)$ is a linear function $Z\theta_j$, with $\theta_j$ an unknown
$K$-vector.
In that case the pliable lasso model (\ref{eqn:plasso}) can be written as
\begin{eqnarray}
\hat y&=&\beta_0 \one+Z\theta_0+ \sum_{j=1}^p X_j (\beta_j \one+  Z\theta_j )\cr
   &=&\beta_0 \one+Z\theta_0+ X\beta+  \sum_{j=1}^p (X_j \circ Z) \theta_j 
\label{eqn:plasso2}
\end{eqnarray}
with $(X_j \circ Z)$ denoting the $N\times K$ matrix  formed by multiplying each column of $Z$ componentwise
by  the column vector $X_j$.
We use the following objective function for this problem
\begin{equation}
J(\beta_0, \theta_0, \beta, \Theta)=\frac{1}{2N} \sum( y_i-\hat y_i)^2   + (1-\alpha) \lambda\sum_{j=1}^p \Bigl((||(\beta_j,\theta_j)||_2+
||\theta_j||_2 \Bigr)  +\alpha \lambda\sum_{j,k} |\theta_{jk}|_1.
\label{eqn:obj}
\end{equation}
Here $\Theta$ is a $p\times K$ matrix of parameters with $j$th row $\theta_j$ and individual entries  $\theta_{jk}$,
The quantities  $\lambda$ and $\gamma$ are adjustable tuning parameters.
The first term in the penalty enforces the hierarchy constraint (\ref{eqn:hier}) while the second term gives sparsity to the individual
components of $\theta_j$. The parameter $\alpha$ controls the relative weight on the two penalties.
Like the usual lasso, for fixed $\alpha$, we obtain a path of solutions indexed by the  tuning parameter $\lambda$.
In fact,  as $\alpha \rightarrow 1$ (but is not equal to 1), the $\hat\theta_j$ are shrunk   to zero and the problem  yields solutions approaching
those of the usual lasso.  Note that the  parameters $\beta_0$ and $\theta_0$ are left unpenalized.

\begin{figure}
\begin{center}
\includegraphics[width=3.25in]{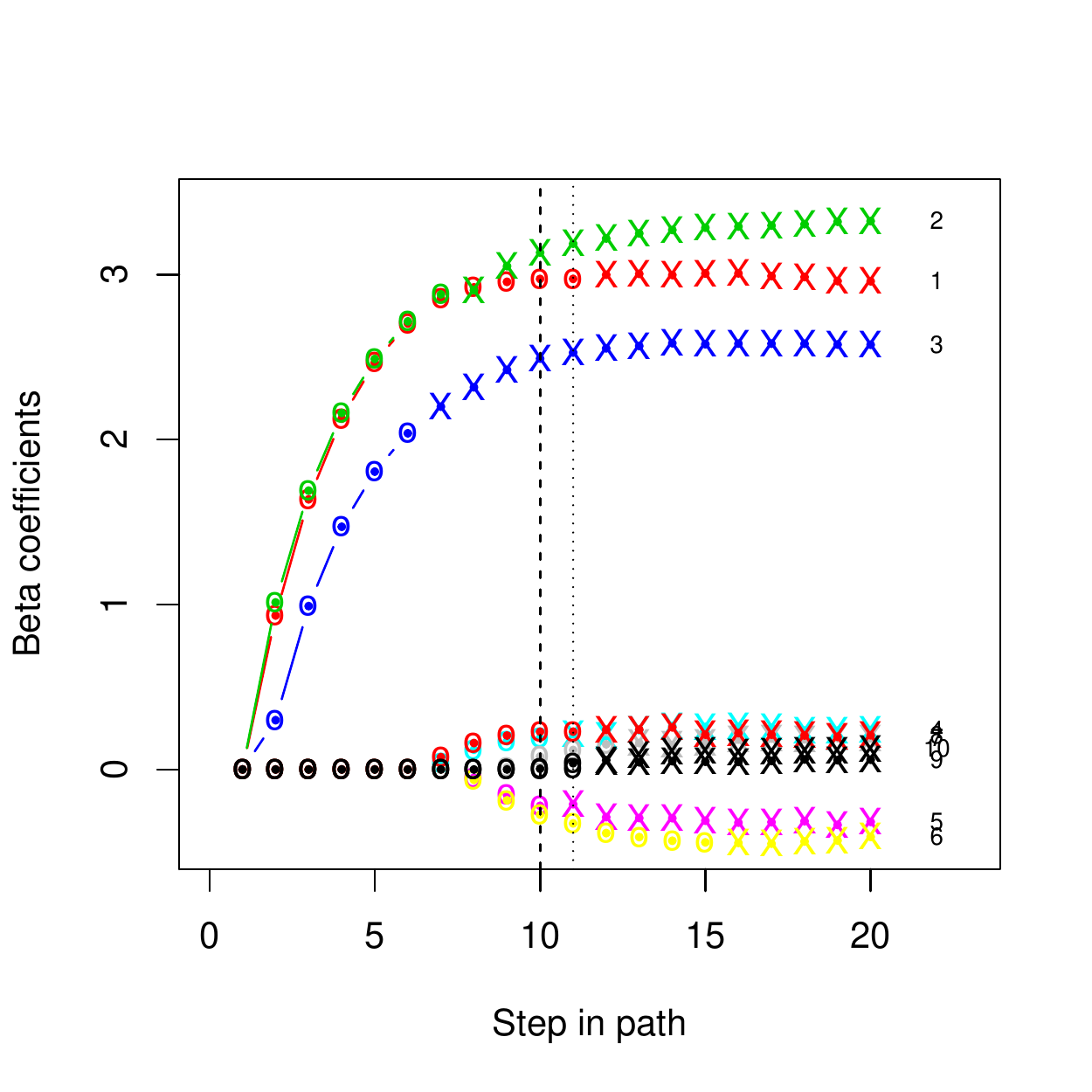}
\end{center}
\caption{\em Solution path for Example 1. The ``X''  symbols indicates that the model has entered a modifying term $Z\theta_j$.
The vertical dotted and broken lines show the model choice corresponding to minimum cross-validation and test error, respectively.}
\label{fig:fig1}
\end{figure}

 Before we give more details, let's see an example. 
 With $N=100, p=20$ and standard normal predictors, we generated
 data from the model
 \begin{equation}
 y=3x_1+ 2x_2+3x_2\cdot I(z=0) + x_3+3x_3 \cdot I(z=1)+0.3\epsilon,
 \end{equation}
 Here $z$ is a single Bernoulli random  variable
and  $\epsilon \sim N(0,1)$.
 Figure  \ref{fig:fig1}  shows the path of solutions as we vary $\lambda$.
The vertical dotted and broken lines show the model choice corresponding to minimum cross-validation and test error, respectively. 
At these values,  see that the model has correctly entered $X_1$ in the usual linear fashion and modifying terms for $X_2$ and $X_3$.
It has also entered (incorrectly) small coefficients for some other predictors.
\subsection{Related work}
There are many connections between the proposal of this paper and previously published work.
There is a close  relationship to the class of varying coefficient models  of  \cite{CH92} and \cite{HT93a}:
the pliable lasso falls into this class, adding the notions of sparsity, hierarchy and scalability to the original proposals.
Our proposed model is also a high-dimensional interaction model: recent work in this area includes  \cite{Zhao2009}, \cite{bach2012}, \cite{BTT2013}, \cite{LH2014},
\cite{Pashova2016} , \cite{she2016},   \cite{family} and  \cite{yan2017}.  The use of the pliable lasso when $Z$ is unknown in the test set relates to the  idea of
customized training   \citep{powers2015}. This is discussed in the next section.
The pliable lasso uses the penalties proposed in the  group lasso   \citep{YL2007} and the sparse group lasso  \citep{simon2013}. 

Unlike interaction models which treat variables all in the same manner,
the pliable lasso is  asymmetric in the way that the variables $X$ and $Z$ are treated.
The $X$ variables are the main predictors in the model, and the modifying terms $\Theta_{jk}$
can only be nonzero if $\beta_j$ is non-zero. But the converse is not true. This is in contrast to other hierarchical interaction models for a set of predictors $X_1 ,X_2 \ldots X_p$: these use either weak hierarchy--- the interaction between $X_j$ and $X_k $ is considered if at least one of their main effects is in the model---, 
or
strong hierarchy, where both main effects have to be in the model. Examples are \cite{BTT2013} and \cite{LH2014}.

This latter procedure of  \cite{LH2014}  (``Glinternet'') uses  a latent overlapping group lasso whereas the pliable lasso uses a group lasso with overlapping groups.
This distinction is made clear in Table 1 of  \cite{yan2017}.
Glinternet models interactions  between all pairs of variables, imposing strong hierarchy on the search: that is, an interaction
can appear only if {\em both} main effects are in the model.  Furthermore, the corresponding R package
allows one to specify ``interactionCandidates'' :  interactions are only considered 
          between  these variables and all other variables. To apply this procedure to our setting, we can take
the set of variables to be $(X,Z)$  and then restrict interactions to the $Z$ variables. In this special case, this differs
from the pliable lasso in the use of strong vs. weak hierarchy, and the fact that the pliable lasso interacts each $X_j$ with the entire vector $Z$. We compare Glinternet  to the pliable lasso
in the simulation examples in the next section.

\subsection{Different use cases}
An attractive property of the pliable lasso is its versatility.  Table \ref{tab:one} summarizes the different possibilities.

\begin{table}
\begin{center}
\begin{tabular}{|llll|}
\hline
Scenario & Training Set & Test Set & Examples\\
\hline
Known-Known& $Z$ known & $Z$ known & gender, age, $Z\subseteq X$\\
Known-Unknown & $Z$ known & $Z$ learned & time, patient ID\\
Unknown-Unknown& $Z$ learned & $Z$ learned & clusters from $X$\\
\hline
\end{tabular}
\end{center}
\caption{\em Different possibilities for the modifying variables $Z$.}
\label{tab:one}
\end{table}

Scenario  ``Known-Known''  is the simplest:  we fit the pliable lasso model   and apply it directly to the test set. One possibility is $Z\subseteq X$, that is, we take $Z$ equal to some (usually small) subset of the $X$ variables. This is illustrated in the HIV example Section \ref{sec:real}.
In Scenario ``Known-Unknown'', we fit our model, but for prediction in the test set, we need to estimate the values of $z$ for the test set. For this purpose we apply a  separate
supervised learning  procedure to predict $Z$ from $x$  in the training set, and then use the predicted values $\hat Z$ when applying the pliable lasso to the test set.
This procedure may also use other information besides $X$, if available, to predict $Z$. In our example we use either  a random forest or multinomial-lasso classifier.

In Scenario ``Unknown-Unknown'' we can form clusters  based on $X$ (or any other available measurements) using either training data, or the combined training and test data.
This scenario and also the scenario ``Known-Unknown''  when $Z$ is a  categorical variable, relates closely to the {\em customized training}  proposal of  \cite{powers2015}. In that work, clusters are estimated from the training data, or the combined training and test sets, and then a local lasso model are fit to each of the clusters. In Section \ref{sec:unknownZ} we experiment with a different approach,
estimating a linear contrast $Z=X\gamma$  from the data itself.
 
 Here is an outline of the  rest of this paper. Section \ref{sec:opt} describes our optimization strategy, while Section \ref{sec:sim} discusses a  simulation study.
 Real data examples are explored in Section \ref{sec:real}.
 Application of the pliable lasso to the  estimation of heterogenous treatment effects is discussed in Section \ref{sec:hetero}. Degrees of freedom of the fitted model
 is discussed in Section \ref{sec:df}, while the setting of unknown $Z$ is covered in Section \ref{sec:unknownZ}. Finally, we discuss extensions of the model in Section \ref{sec:further}.

 \section{Optimization of the objective function.}
 \label{sec:opt}

The objective function (\ref{eqn:obj}) is convex, and for its minimization
we use  a blockwise cyclical coordinate descent procedure, over a path of decreasing $\lambda$ values
using the previous solutions as warm starts.
 The problem has some attractive structure that simplifies the computation:
in particular, fixing the other predictors, we have derived explicit conditions for determining if $(\hat\beta_j, \hat\theta_j)$ is nonzero and
if it is nonzero,  whether the  $\hat\theta_j$ component is nonzero. 
If both are nonzero, we use a  generalized gradient descent procedure to determine both parameters.
This strategy makes for a fast algorithm, since we can cycle through the predictors and only do the costlier computation (generalized gradient descent)
when needed.

The generic form of the algorithm is shown in  Algorithm \ref{alg1}.

\begin{algorithm}
\caption{ Algorithm for the Pliable Lasso}

\begin{description}
\item For a decreasing path  of $\lambda$  values:
\begin{description}
\item Repeat until convergence:
\begin{enumerate}
\item Compute $\hat\beta_0$ and $\hat\theta_0$  from  the  least squares regression of the current residual on $Z$.
\item For predictor  $k=1,2,\ldots p,1,2,\ldots $ 
\begin{description}
\item{(a)}  Check an (explicit) condition for $(\hat\beta_j, \hat\theta_j)=0$. if zero, skip to next $k$
\item{(b)}  Otherwise,  compute $\hat\beta_j$ using soft-thresholding and then check the condition for  $\hat\theta_j=0$. If zero, fix update $\hat\beta_j$ and then move to next $j$
\item{(c)} Otherwise, if both ($\hat\beta_j, \hat\theta_j$) are nonzero:
use a generalized gradient procedure to find  $(\hat\beta_j,\hat\theta_j)$.
  \end{description}
\end{enumerate}
\end{description}
\end{description}
\label{alg1}
\end{algorithm}

\bigskip
The details of this procedure are given in the Appendix. 
As shown there, the condition used in Step 1 of the algorithm is
\begin{equation}
|X_j^Tr_{(-j)}/N| \leq (1-\alpha)\lambda\; {\rm  and} \;
||S(W_j^Tr_{(-j)}/N,\alpha\lambda)||_2 \leq 2(1-\alpha)\lambda
\label{condtemp}
\end{equation}
where $r_{(-j)}$ is the partial residual with the fit for the $j$th predictor removed,
 $W_j=X_j\circ  Z$ (elementwise multiplication in each column)
and $S(x, t)={\rm sign}(x)(|x|-t)_+$ , the soft-threshold operator.
We note the similarity to the standard coordinate descent
procedure for the lasso (see e.g. \cite{FHT2010}).  Specifically, without the modifiers $Z$, the second condition
disappears and the first expression is exactly the zeroness condition in the coordinate descent procedure for the lasso.
\medskip

{\bf Remark A.} We have included a factor $N$ in the denominator of the first term of the objective  function in (\ref{eqn:obj}), to match the
parameterization used in the {\tt glmnet} program \citep{FHT2010} .
 \medskip
 
 {\bf Remark B.} The solutions to the optimization in (\ref{eqn:obj})  depend on the scaling to the $X$ and $Z$ variables.
 By default, we standardize each set to have zero mean and unit variance.
 \medskip
 
 {\bf Remark C.} 
In the coordinate descent procedure for the lasso procedure  of \cite{FHT2010}), using ``naive'' updates, the coefficient for each predictor
can be checked and updated in $O(N)$ operations. In the current algorithm, this cost increases to $O(N\cdot K$) operations.

 \section{A simulation comparison}
 \label{sec:sim}
 In this example we took $n=100, p=50$, and standard Gaussian independent predictors.
 The response was generated as
 \begin{eqnarray}
 y&=&\mu(x)+ 0.5 \cdot\epsilon;\;\;\epsilon \sim N(0,1)\cr
 \mu(x)&=&x_1\beta_1+x_2\beta_2+  x_3(\beta_3e+ 2 z_1)  +x_4\beta_4( e-2z_2);\; \beta = (2,-2,2,2,0,\cdots) 
\label{eqn:simmodel}
\end{eqnarray}
with $e=(1,1,1\ldots 1)^T$. The modifying variables $Z$ were drawn from a simple Bernoulli distribution with equal probabilities.
The signal to noise ratio was about 2.
 We applied the standard lasso, (using the R package {\tt glmnet}),  gradient boosting machines (using the R package {\tt gbm}) ,  Glinternet   (using the R package {\tt glinternet}), and the pliable lasso over 20 simulations.
 The sample size $N$, $p$= number of $X$ variables, $K$=number of $Z$ variables are shown at the
 top of each panel. 
 
 Different variations of the problem are used in each panel, with details given in  the figure caption. In every case the pliable lasso does best.
 As noted in the caption, all methods were run with default settings; while we didn't tune the pliable lasso in any special way, it is possible
 that GBM and glinternet could perform better using different parameter settings.
 Of course GBM fits a much more general model than the pliable lasso and should  perform  better in more general settings, but it is reassuring than the pliable  lasso
 does well across this range of problems.
 
  \begin{figure}
 \includegraphics[width=6in]{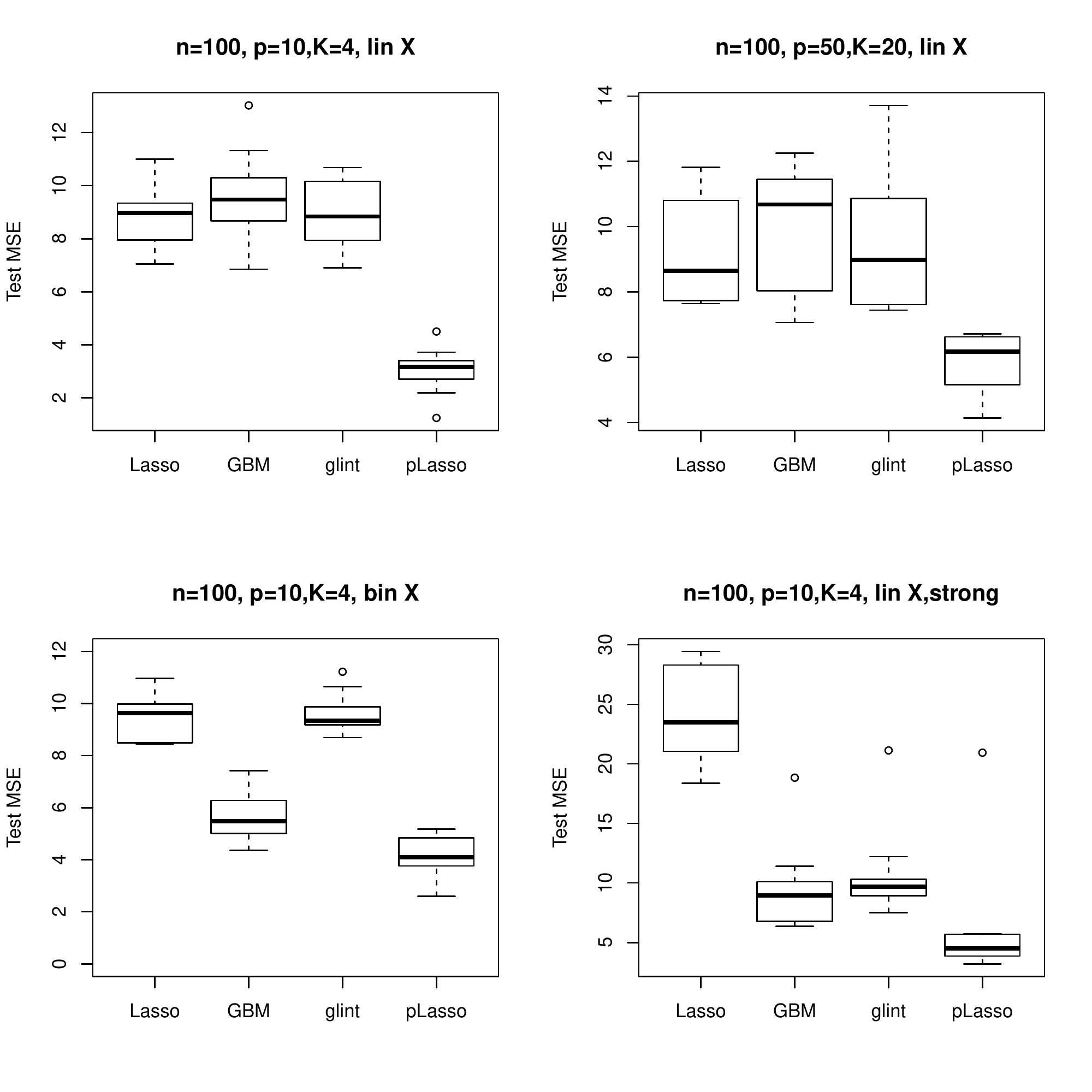}
 \caption{\em Results from simulation experiment using model (\ref{eqn:simmodel}) and it variants.
 The methods are lasso, GBM, glinternet  and the pliable lasso.
 The lasso, GBM and glinternet were given both  $X$ and $Z$ as predictors.
 GBM used single split trees (stumps), which seems appropriate since the true interactions are first-order. In glinternet we only allowed  interactions
 between $Z$ and the other predictors $X$.
 Otherwise, all methods were run with default settings and  each used cross-validation to choose its tuning parameters.
 The top  left panel uses model (\ref{eqn:simmodel}),with $N=100, p=10$, while top top right panels has $N=100, p=50$.
 In In the bottom left panel, we used $I(X_j>0)$ in place of each $X_j$, to give a potential advantage to GBM. In the bottom
 right we added strong main effect terms in $Z$, to aid the strong hierarchy strategy of glinternet.}
 \label{fig:sim1errorL}
\end{figure}

 \section{Real data examples}
 \label{sec:real}
 \subsection{A $Z\subseteq X$ example:  HIV mutation data}
 
\begin{figure}
 \begin{center}
 \includegraphics[width=2.75in]{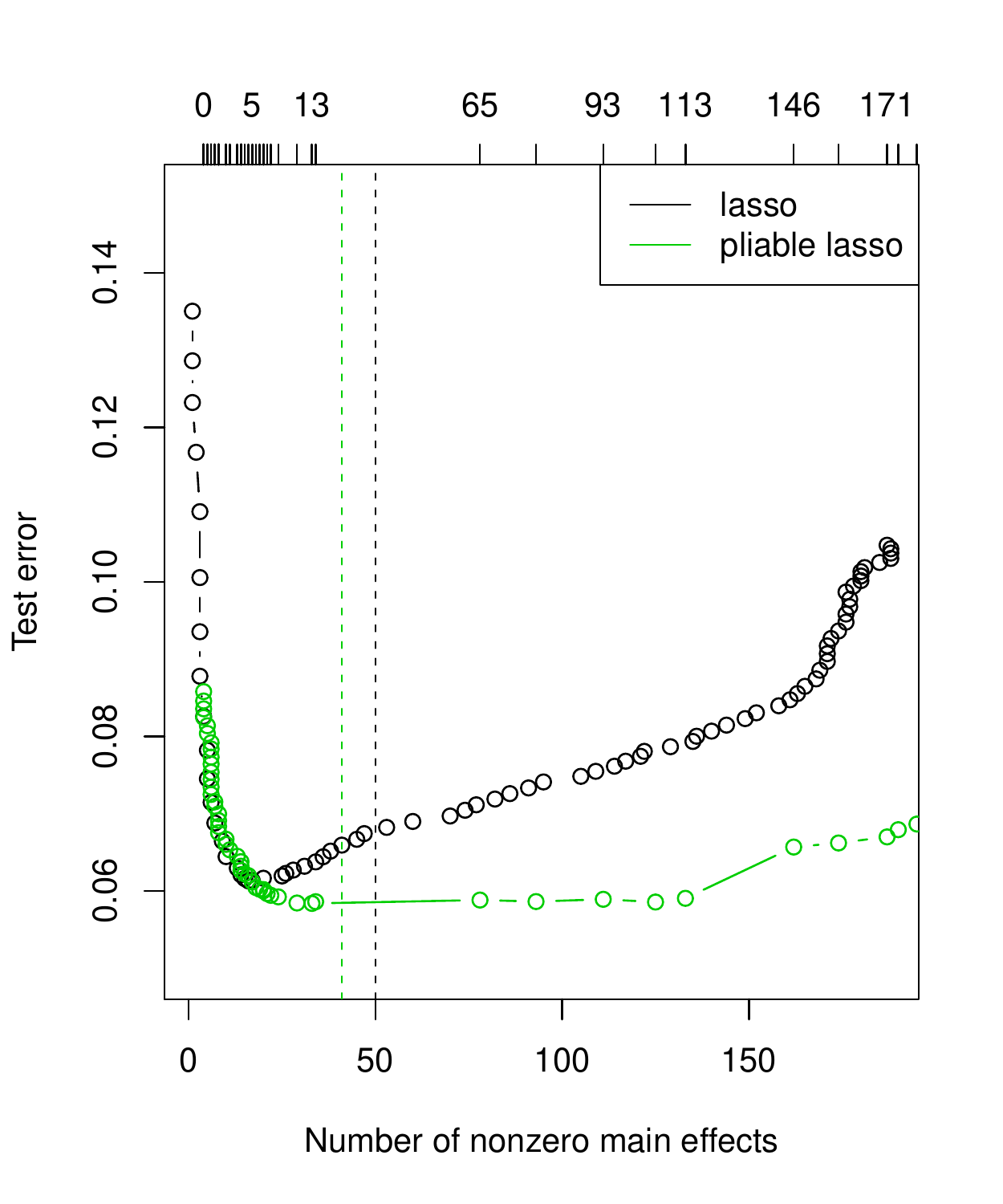}
 \end{center}
 \caption{\em HIV mutation data:  test error curves for the lasso and pliable lasso. The numbers across the top of the plot represent the number of main effects  having interaction terms. The vertical broken lines show the choice of model from cross-validation.}
 \label{fig:nrtiL}
\end{figure}

  \citet{rhee2003} study six nucleotide reverse transcriptase inhibitors (NRTIs) that are used to treat HIV-1. 
  The target of these drugs can become resistant through mutation, and they compare a collection of models for predicting these drug's (log) susceptibility-- a measure of drug resistance, based on the location of mutations.
We focussed on the first drug (3TC), for which there are
 there are $p = 217$ sites and  $N=1057$ samples. We randomly divided the samples into approximately equal-sized training and test sets.
 In this case we chose the modifiying variables $Z$ to be a subset of the $X$ variables:  we used the mutations having training set univariate $z$-scores $\ge 2$ in absolute value.
 Figure \ref{fig:nrtiL} shows the test error curves for the lasso and pliable lasso. We see that pliable lasso achieves somewhat lower test error than the lasso.
 The resulting model is easy to understand, involving some main effect mutations  and some pairwise interactions between pairs of mutations.
 
 \subsection{Skin cancer proteomics data}
The data in this example come from DESI mass spectometry measurements from tissues of patients with skin cancer, from collaborators here at Stanford. 
There are 16 patients in the training set, contributing  with a total of 17,053 measurements (there are many image pixels per patient) and 
2733 proteomic features. Each pixel is labelled as normal or cancer, with most patient samples
consisting entirely of normal or cancer pixels. The test set had 7963 measurements from 9 patients.  For computational ease we  took a random sample of 1000 measurements from the training set and chose the 1000 features $X_j$ having largest 
univariate scores in the training set. We have used the lasso with success on similar data with other cancers, see e.g \cite{eberlin2014}.

Here we took $Z$ to be the patient ID. This is an example of the ``known-unknown'' scenario of Table \ref{tab:one}:
the patient IDs in the test set are different from those in the training set. Hence we fit  a separate (multinomial-lasso)
classifier in the training set, and use it to ``predict'' $Z$ in the test set. In other words, we find the training set
patient most similar to a given  test set patient, and use his/her patient ID for the prediction. To further
enhance accuracy, we further filtered the $X_j$ variables, keeping the ones with  univariate scores in the top 1000
for the outcome (as mentioned above ),  but also required that they had scores in the top 1000 with respect to  their variation between patients. 
This encourages the procedure to build a model on the training set that involves more  predictable $Z$  variables,
so that it will be more effective when applied to the test set.  The second filtering left 437 features in the training set for the pliable lasso.  The 16-class classifier for predicting patient ID in the training set had an  cross-validation error rate  of about
40\%.

\begin{figure}
 \begin{center}
  \includegraphics[width=3.5in]{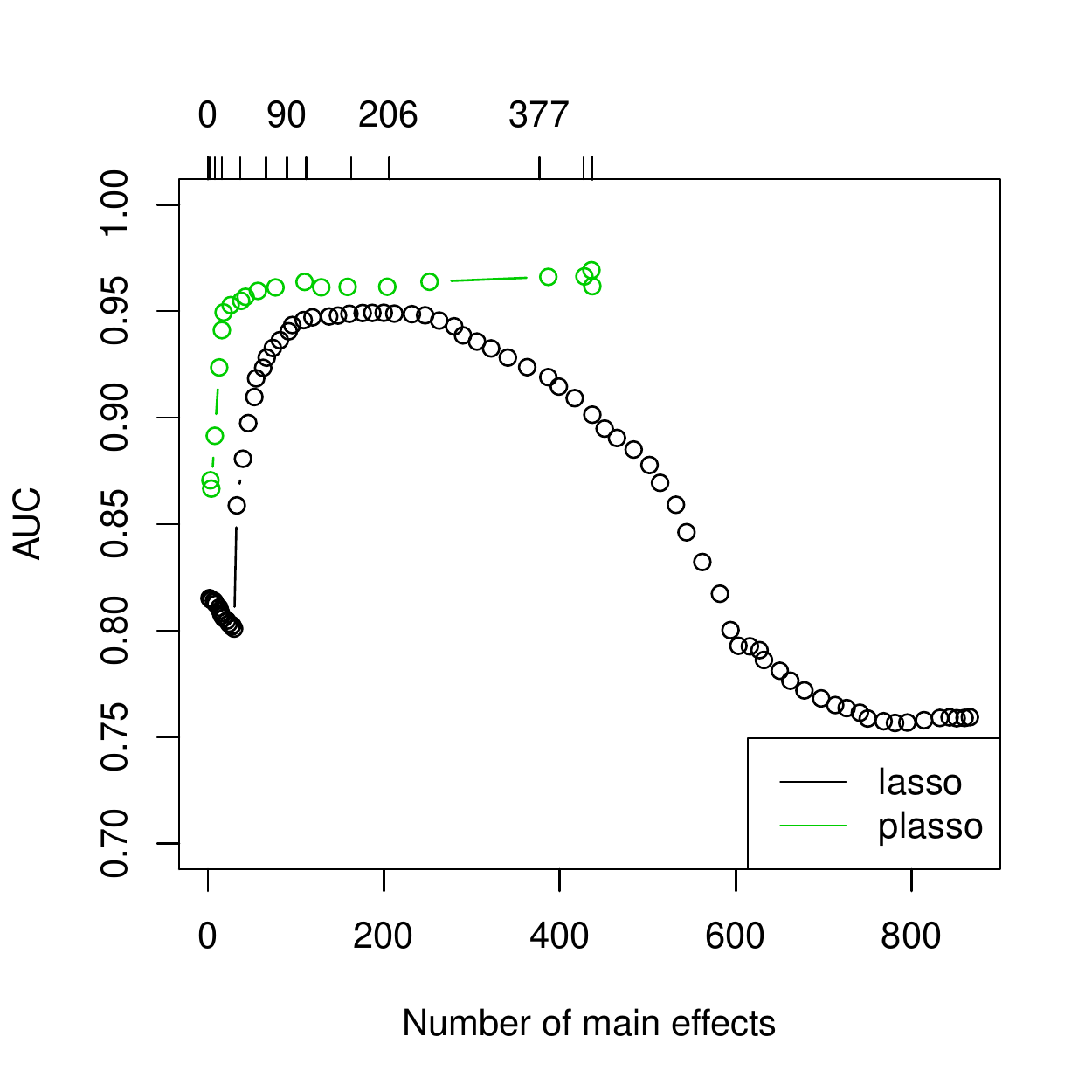}
 \end{center}
 \caption{\em Skin cancer data: test set area under the curve for the lasso and pliable lasso, for various model sizes.}
\label{fig:auc-skinL}
\end{figure}

The resulting test set AUCs  for the lasso and pliable lasso are shown in Figure \ref{fig:auc-skinL}.
While the improvement shown by  the pliable lasso over the lasso might seem small, it is actually quite
significant: from a best rate of 94\% for the lasso to about 97\% for the pliable lasso.

\subsection{Forecasting example- predicting stock returns}

\begin{figure}
\centerline{
\includegraphics[width=3.5in]{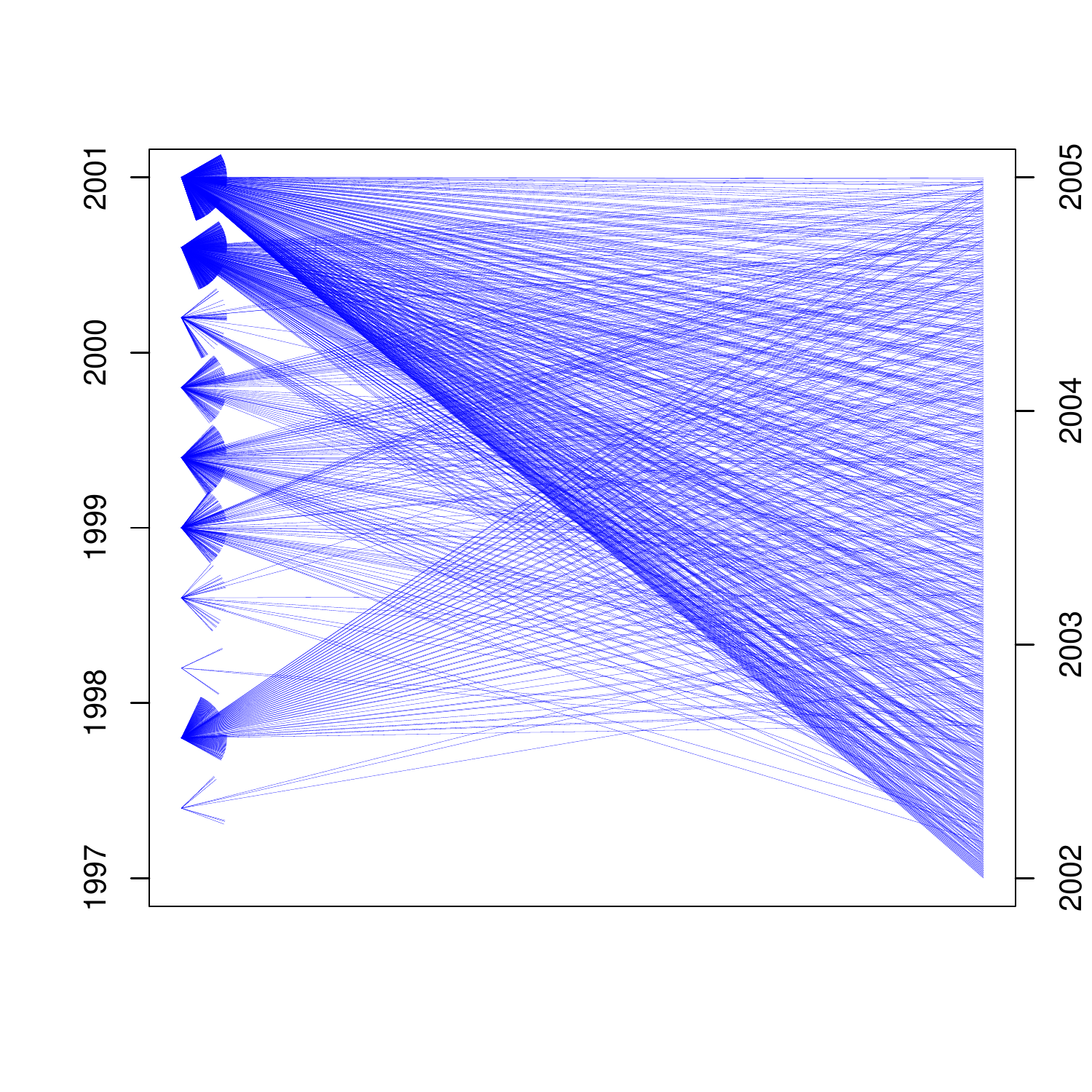}
}
\caption{\em Stock return data: each time on the right is classified as being most  similar to the training time period
indicated on the left. }
\label{fig:arrows}
\end{figure}

Here is a another example of the ``known-unknown'' scenario of Table \ref{tab:one}.
The data are 21 day returns for a stock.
We make daily predictions  by  fitting a lasso using 33 available signals as features.
The training and test sets cover
the periods 1997-2001 and  2002-2005 respectively.
The base model uses a common lasso model fit to all of the training data.
For  the pliable lasso, we divided the training period into 10 equal time periods, and set   $Z$ equal  to
the resulting ten category variable.  We then 
built a random forest classifier to predict the most similar  training period  for each test date.
Figure \ref{fig:arrows} depicts the classification of each test date.

The correlations of each predicted return with the actual return in the test set are shown in 
Figure 6.
The pliable lasso achieves an increase of nearly 2\% over the lasso, which could be quite significant practically.

\begin{figure}
\begin{center}
\includegraphics[width=3.5in]{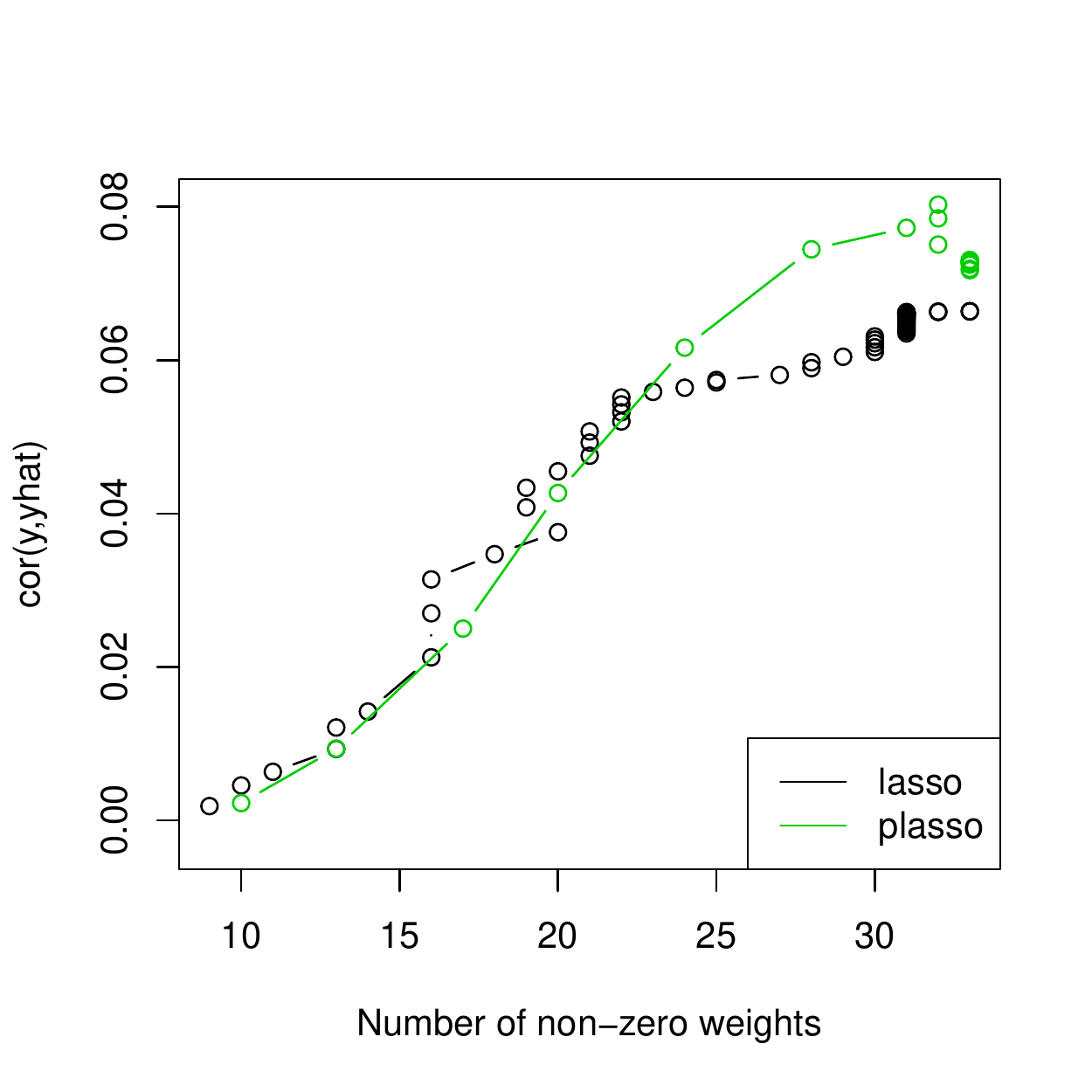}
 \end{center}
 \label{fig:clintcor}
 \caption{\em Stock return data: correlations between predicted and actual test period returns, for the lasso and pliable lasso.}
\end{figure}
\subsection{Example: States crime data}
\label{sec:crime}
\begin{figure}
\centerline{\includegraphics[width=4in]{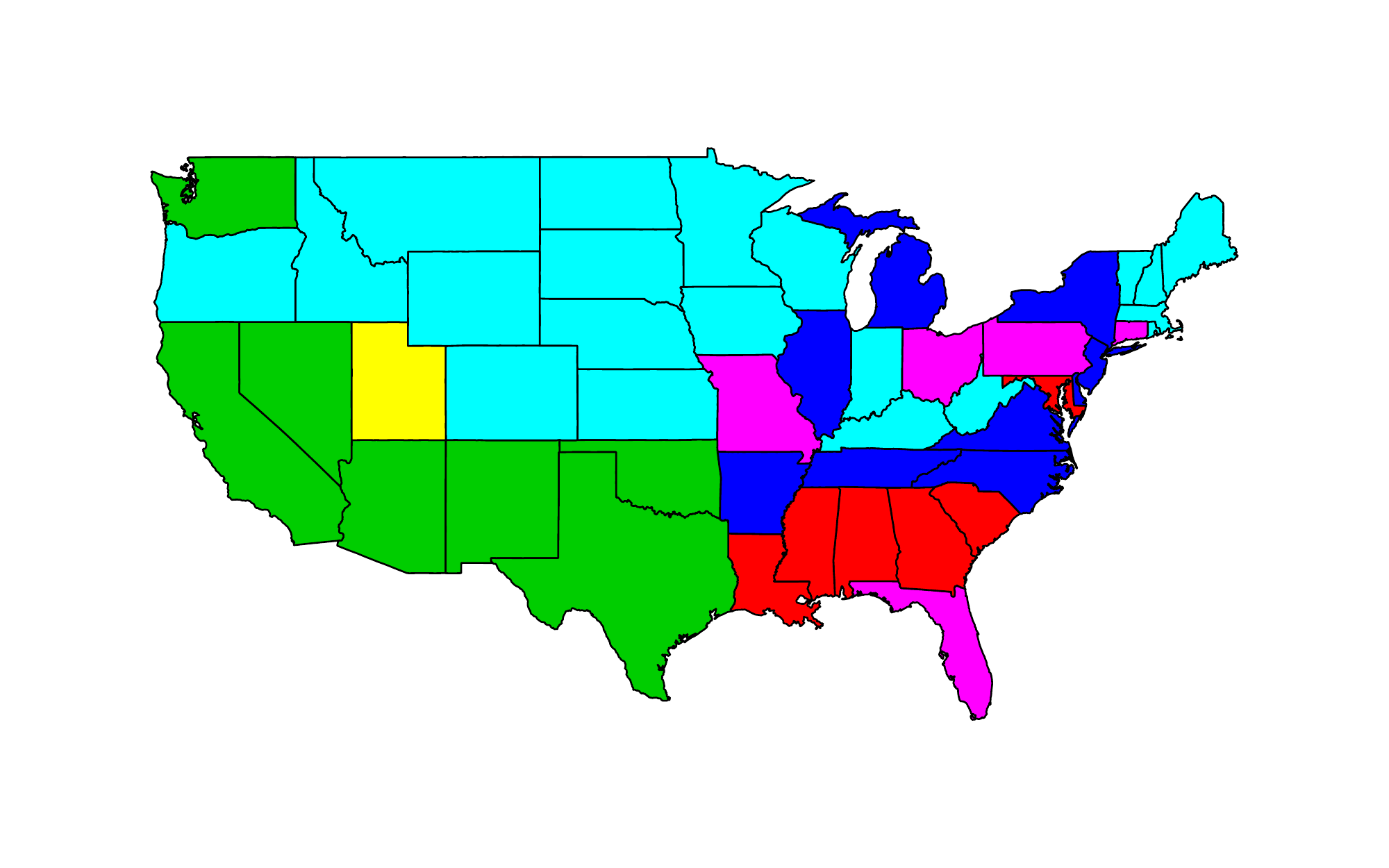}}
\caption{\em Crime data: grouping of states based on clustering of  demographic variables.}
\label{fig:usagroups}
\end{figure}

\begin{figure}
\centerline{\includegraphics[width=3.75in]{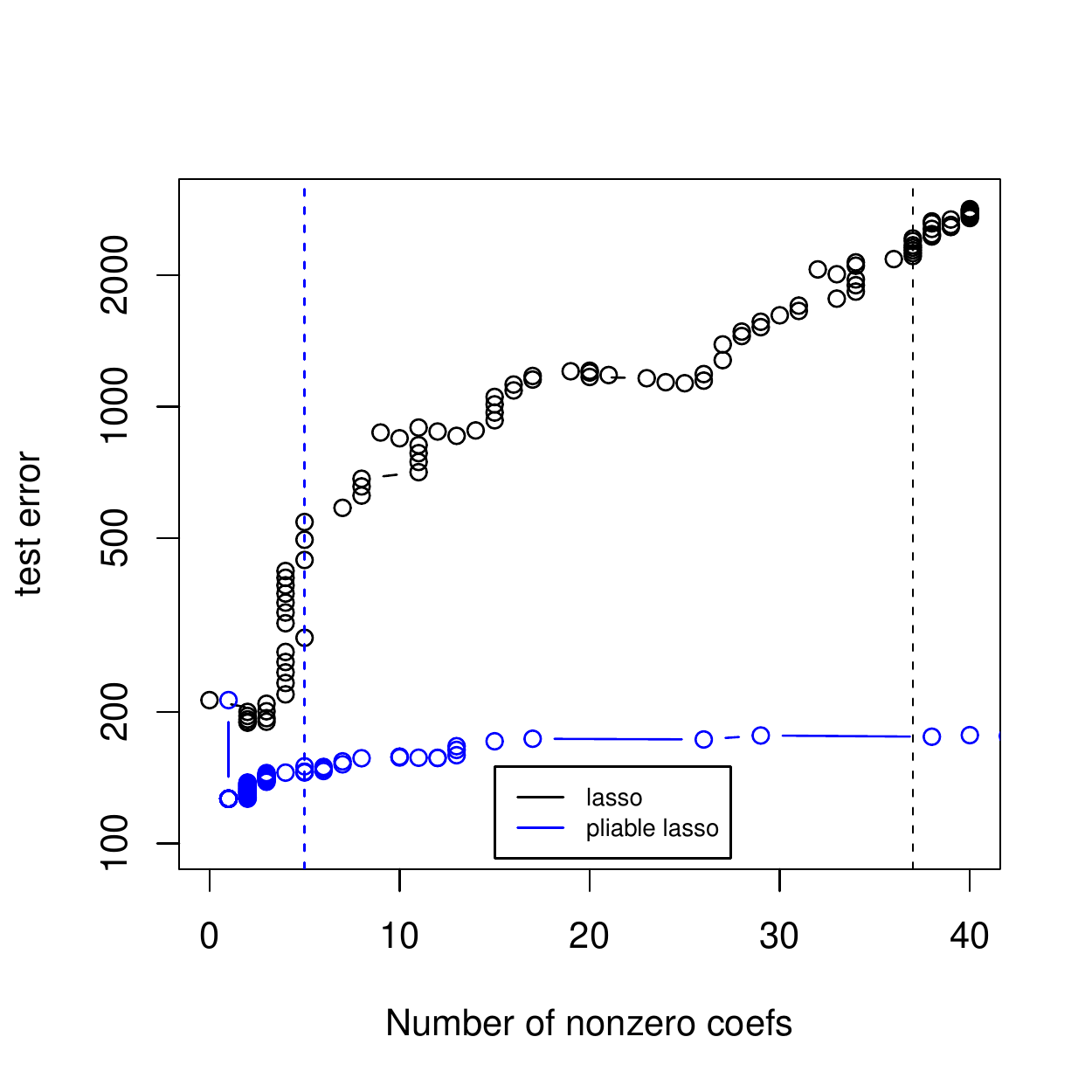}}
\caption{\em Test error rates for the crime data. The choice of model size via cross-validation is indicated by the vertical broken lines.}
\label{fig:crimeerr}
\end{figure}

\begin{figure}
\centerline{\includegraphics[width=3.25in]{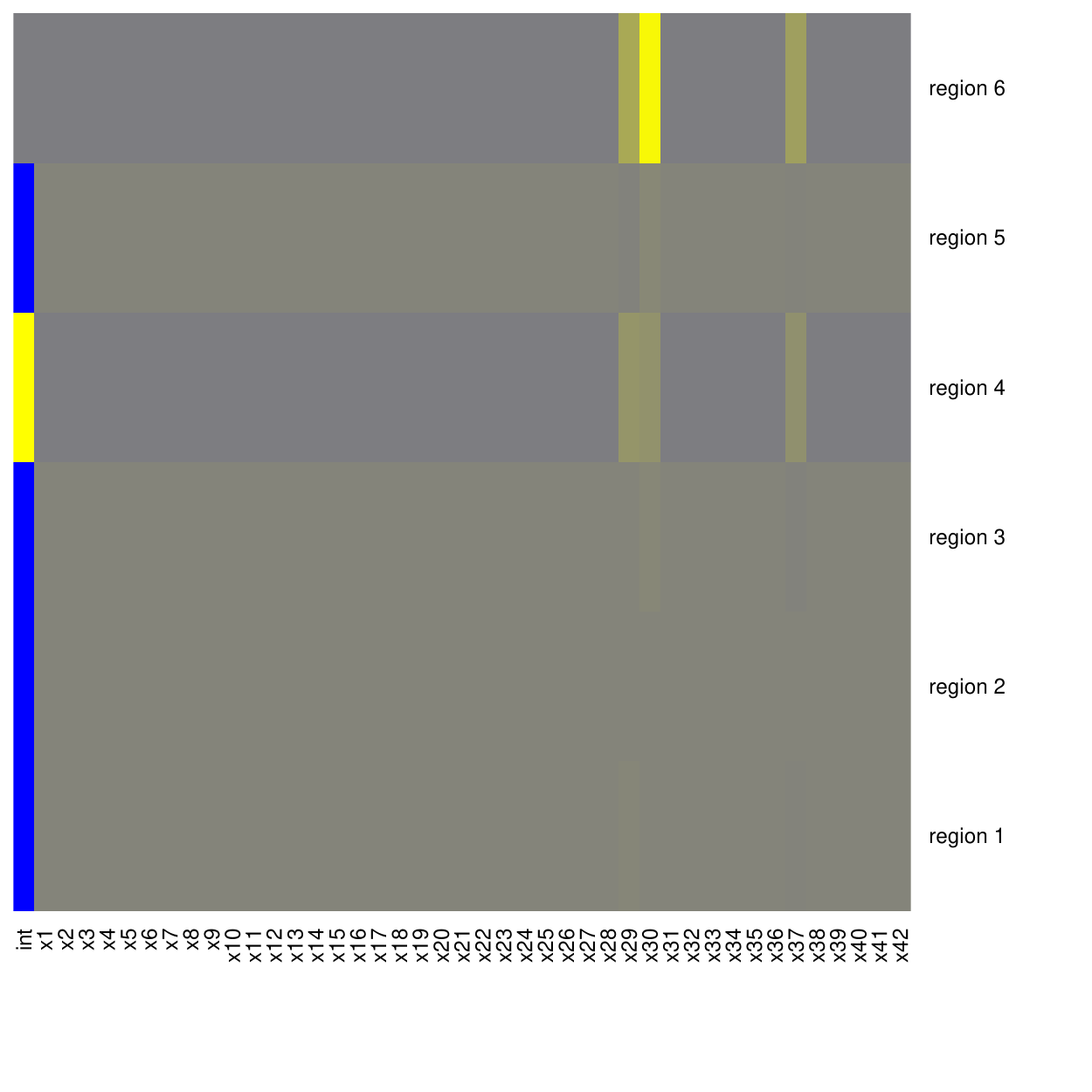}}
\caption{\em Crime data: estimated modifying coefficients $\hat\theta_{jk}$  for predictor $j$ (column of heatmap)
and region $k$ (row of heatmap). Referring to colors of Figure \ref{fig:usagroups}, numbering is
red=region 1,  green=region 2,  dark blue=region 3,  light blue=region 4, purple=region 5, yellow=region 6.}
\label{fig:heatcrime}
\end{figure}

In the section we analyze a public dataset on  yearly state crime statistics,  from 1977-2014,  obtained from colleague John Donahue of Stanford law school.
The outcome is violent crime rate  and the predictors are 42 demographic variables. We created training and test sets by dividing the data into
two approximately equal time periods.
We first applied hierarchical clustering to the predictors, yielding the six regions depicted by  Figure \ref{fig:usagroups}. 
We fit a pliable lasso model with $z$ equal to the six-category location.
 The test set error rates for  the lasso and pliable lasso
are shown in Figure \ref{fig:crimeerr}. The choice of model size via cross-validation is indicated by the vertical broken lines in the plot. We used standard cross-validation, ignoring time ordering, and this produced a poor model size estimate for the lasso. Perhaps not surprisingly, the  pliable lasso is able to use  the region similarity
to improved prediction. A heatmap of the  modifying coefficients $\hat\theta_{jk}$ is shown in Figure \ref{fig:heatcrime}.
The unpenalized intercept $\hat\theta_0$ appears at the left and has the most effect. Two or three demographic
predictors also vary across region. Note that the resulting model is somewhere in between a common model for all
regions and separate linear model for each region.

\section{Estimation of heterogenous treatment effects}
\label{sec:hetero}
The estimation of heterogenous treatment is a ``hot'' area of research, especially promising for the area of personalized medicine.
The idea there is to find subsets of patients who will benefit from a specific treatment regime. This problem is especially challenging
with high-dimensional features and observational data, that is, non-randomized treatment assignments. The small effect sizes seen in many real datasets makes the problem even more challenging. A review of recent work in this area is given by \cite{powers2017}.  Here we  briefly explore  the application of the pliable lasso to this setting.
Our data has the form $(X_i, W_i, Y_i)$ where $X_i$ is a vector of covariates, $Y_i$ a real-valued response and $W_i \in \{0,1\}$ a
treatment assignment. We assume that the treatment is randomly assigned with equal probability and make  the usual
unconfoundedness assumption. Extension to non-randomized studies via propensity scores is possible, but will not be explored here.

We use the pliable lasso model  (\ref{eqn:plasso2}) with $Z=W$; the treatment effect at $x$ is estimated by
$\hat y(x,W=1)- \hat y(x,W=1)$.
With  50 standard normal predictors,  we generated 100 observations from three different models:
\begin{eqnarray}
(A): \;y &=& x_1+ x_1w + x_2 +2x_3  +2\epsilon;\; \epsilon \sim N(0,1)\cr
(B): \;y &=& x_1(w-0.5)+ x_2 +2x_3  +2\epsilon;\; \epsilon \sim N(0,1)\cr
(C): \;y &=& x_1+ I(x_1>0)w + x_2 +2x_3  +2\epsilon;\; \epsilon \sim N(0,1)
\label{eqn:hte}
\end{eqnarray}
The signal to noise ratio is about 1.5.
Note that scenario A in the ``home court'' for the pliable lasso, with a single linear hierarchical interaction.  Scenario B  has a non-hierarchical interaction, centered  with respect to the main effect. In Scenario C the interaction is hierarchical, but the
predictor $X_1$ is dichotomized at 0.

We compared the pliable lasso approach to  causal forests \citep{wager2015}, a state of the art method for this problem.
It is implemented in the generalized random forest package of \cite{grf}: we used version 0.9.3 with  default settings.
For scenarios A and B, the pliable lasso performs best: the advantage in (B) seems surprising--- perhaps the random imbalance 
in treatment assignment creates a small main effect for $X$. In Scenario C, where the causal forest can split $x_1$
appropriately, it shows smaller variance than the pliable lasso, but with some bias.
Of course the causal forest might perform better with other parameter settings, and is able to model much more general, high-order interactions than the pliable lasso.

\begin{figure}
\centerline{\includegraphics[width=5in]{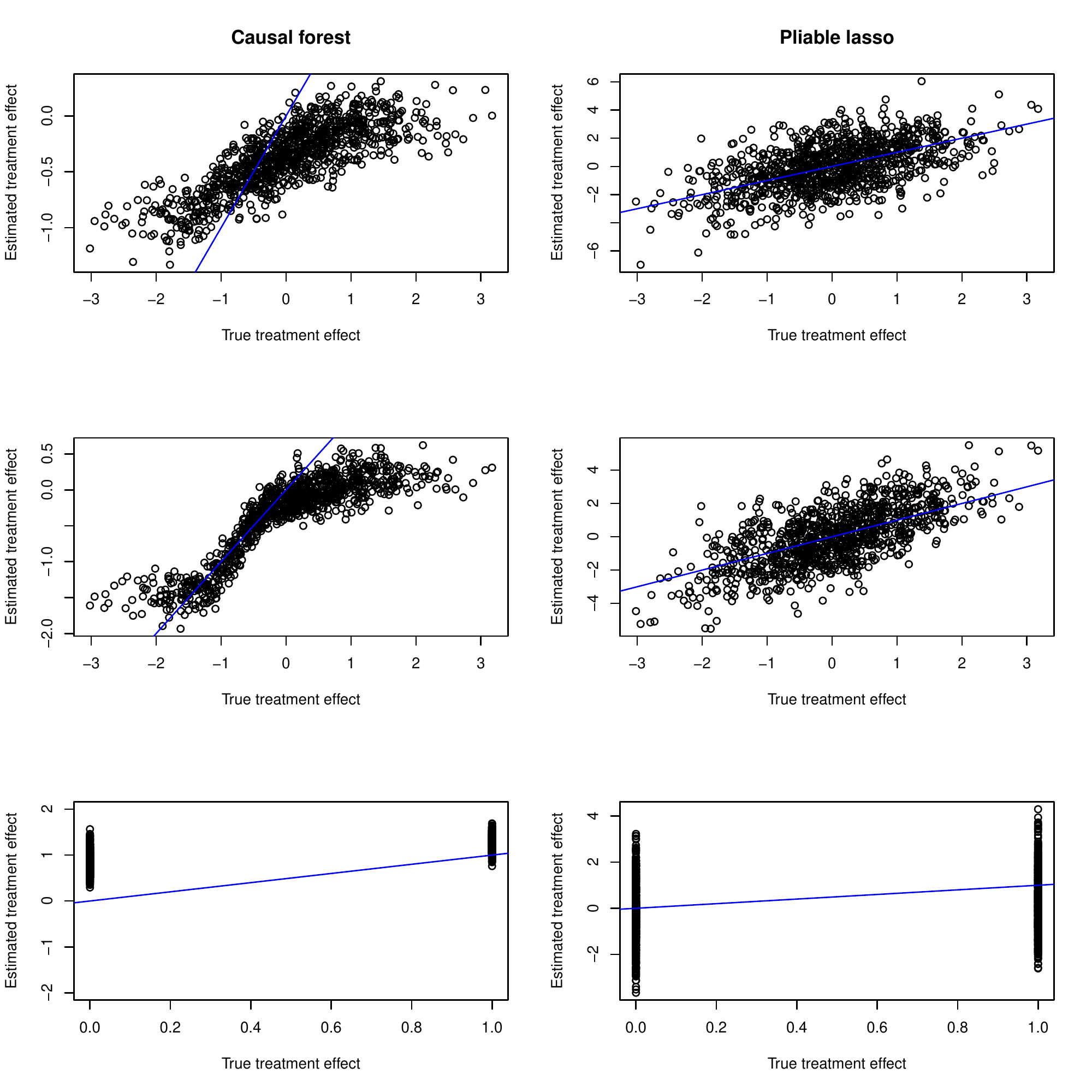}  }
\caption{\em Results for causal forests and pliable lasso applied to the three problems in (\ref{eqn:hte}). The top, middle
and bottom panels correspond to scenarios A, B, and C. The 45 degree line is drawn in blue each panel.}
\label{fig:treff}
\end{figure}

\section{Degrees of freedom of the fit}
\label{sec:df}
Given a vector of response values $y$ and a fit vector $\hat y$, \cite{Ef86} defined the degrees of the fit by

\begin{equation}
{\rm df} (\hat y)\equiv   \frac{\sum_i{\rm Cov}(y_i,\hat y_i)}{\sigma^2}
\label{eqn:cov}
\end{equation}
The power of this definition come from the fact that it can be applied to non-linear, adaptive estimators. 

Now \cite{LARS} shows that if $y\sim N(\mu,I)$, then for the least angle regression procedure   (a method for constructing the lasso path)
after $k$ steps the degrees of freedom equals $k$.
This result was strengthened and generalized in  \citet{ZHT2007} and \cite{lassodf2} to show that for the lasso
the number of non-zero elements in the solution is an unbiased estimated of the degrees of freedom.

\begin{figure}
\includegraphics[width=6in]{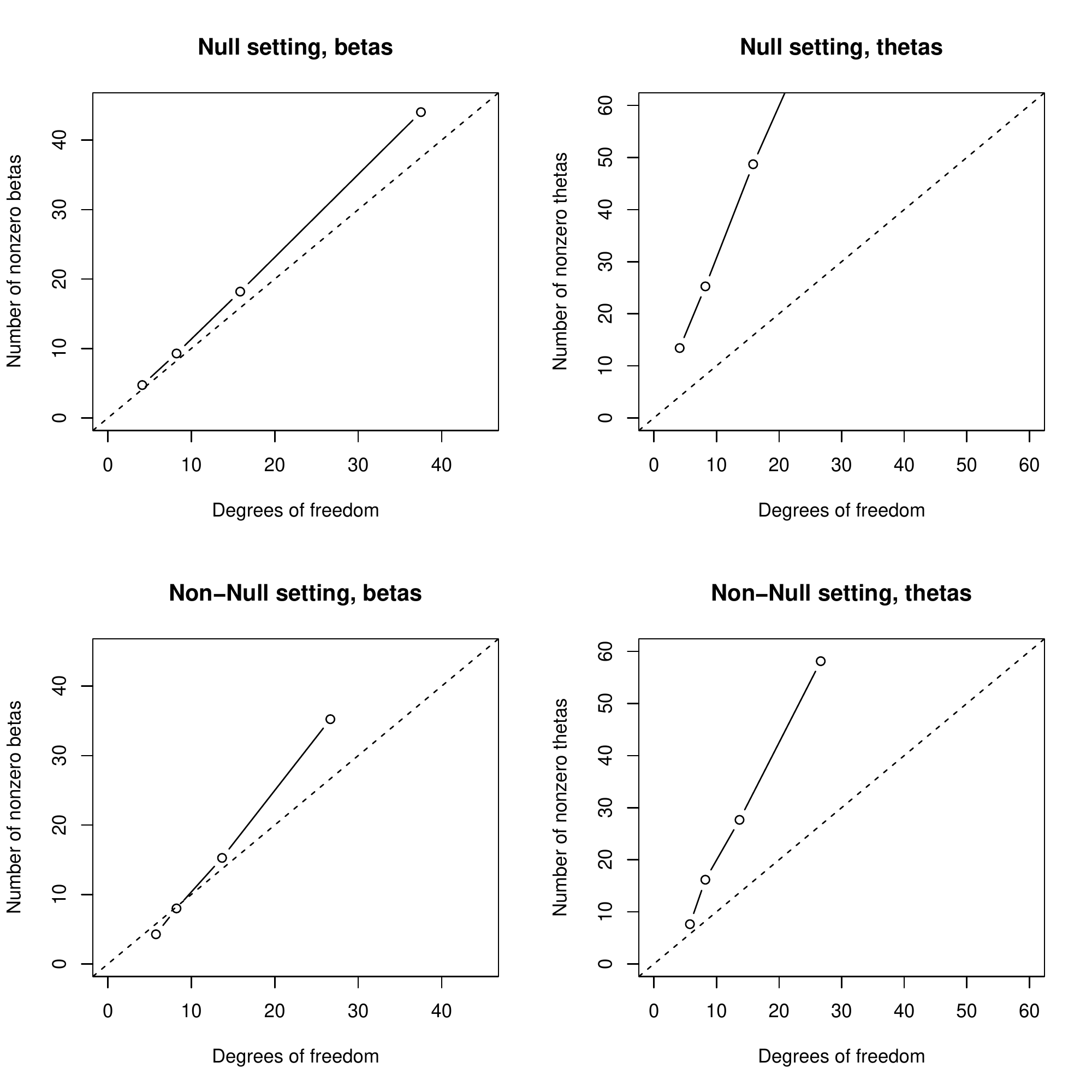}
\caption{\em Estimates of degrees of freedom for the pliable lasso, in the null setting (top panels) and the non-null setting (bottom panels).  The degrees of freedom (horizontal axes) are estimated from the covariance formula, while the number of
non-zero $\beta$ and $\theta$ parameters are shown in the left and right panels.}
\label{fig:dfL}
\end{figure}

Since the pliable lasso is a generalization of the lasso, we ask the question: how many degrees of freedom are spent
in fitting a pliable lasso model with $k$ terms?  In principle, this quantity  may be  analytically tractable but we have not yet succeeded in 
its derivation.
Hence we turn to simulation to shed light on this question. In our setup, we take $N=100, p=5, 10,  20,  50$, and generate standard normal predictors and the outcome
from a null model and a non-null model with error variance one.   The results are in Figure \ref{fig:dfL}.
We used  estimated the covariance in (\ref{eqn:cov}) via the bootstrap, to provide an estimate of  the degrees of freedom (horizontal axes).

We see that the number of non-zero $\hat\beta_j$s provides a rough estimate of the degrees of freedom of the fit  (left panels).
On the other hand, the number of non-zero parameters including the $\hat\theta$s parameters is a gross over-estimate of the degrees of freedom (right panels). This is intuitively reasonable; the hierarchy constraint limits the number of modifying terms, when one does enter,
the coefficients of both main effects and modifiers are shrunken.  It would useful to investigate this rough ``conjecture'' in future work.

\section{ The setting of  unknown $Z$}
\label{sec:unknownZ}
In this section we consider the pliable lasso model  
\begin{equation}
\hat y=\beta_0 \one+Z\theta_0+ X\beta+  \sum_{j=1}^p (X_j \circ Z) \theta_j
\label{eqn:plasso3}
\end{equation}
but now assume that $Z$ is not observed.
For simplicity we assume that $Z$ is a column vector that can be approximated by a linear function of $X$ that is  $Z= X\gamma$, 
 with $\gamma$ an
unknown $p$-vector (extensions to matrix-valued $Z$ are also possible). We estimate $\gamma$ with an $\ell_2$ penalty.
The  objective function for the enlarged problem is
\begin{equation}
J(\beta_0, \theta_0, \beta, \Theta, \gamma)=\frac{1}{2n} \sum( y_i-\hat y_i)^2   + (1-\alpha) \lambda\sum_{j=1}^p \Bigl((||(\beta_j,\theta_j)||_2+
||\theta_j||_2 \Bigr)  +\alpha \lambda\sum_{j,k} |\theta_{jk}|_1 +\frac{\lambda_2}{2} ||\gamma||_2^2.
\label{eqn:obj2}
\end{equation}
This problem is not convex, but is bi-convex--- convex in $\gamma$ with other parameters fixed, and vice-versa.

The two subproblems can be easily solved. With $\hat\gamma$ fixed, we solve the original pliable lasso problem  (\ref{eqn:obj}) with
$Z=X\hat\gamma$. With the other parameters fixed, we write $W=\hat\theta_0 X+\sum_{j=1}^p (X_j\hat\theta_j *X)$
 and  $r=y-\hat\beta_0-X\hat\beta$
and solve
\begin{equation}
{\rm minimize} \frac{1}{2n} \sum( r_i-W\gamma)^2  +\frac{\lambda_2}{2} ||\gamma||_2^2
\end{equation}
This is just a ridge regression  without an intercept. In principal one could alternate these two steps until the procedure hopefully converges.
I
To investigate this procedure, we simulated data with $n=200, p=12$ in two regimes  

$$y=x ^T\beta_z+\sigma\epsilon;  \;\; {\rm with} \; z=0:  \beta_0=(2,2,2,2,0,\ldots 0);  \;\; z=1:  \beta_1=(-2,-2,-2,-2,0,\ldots 0);$$

\begin{figure}
\includegraphics[width=6in]{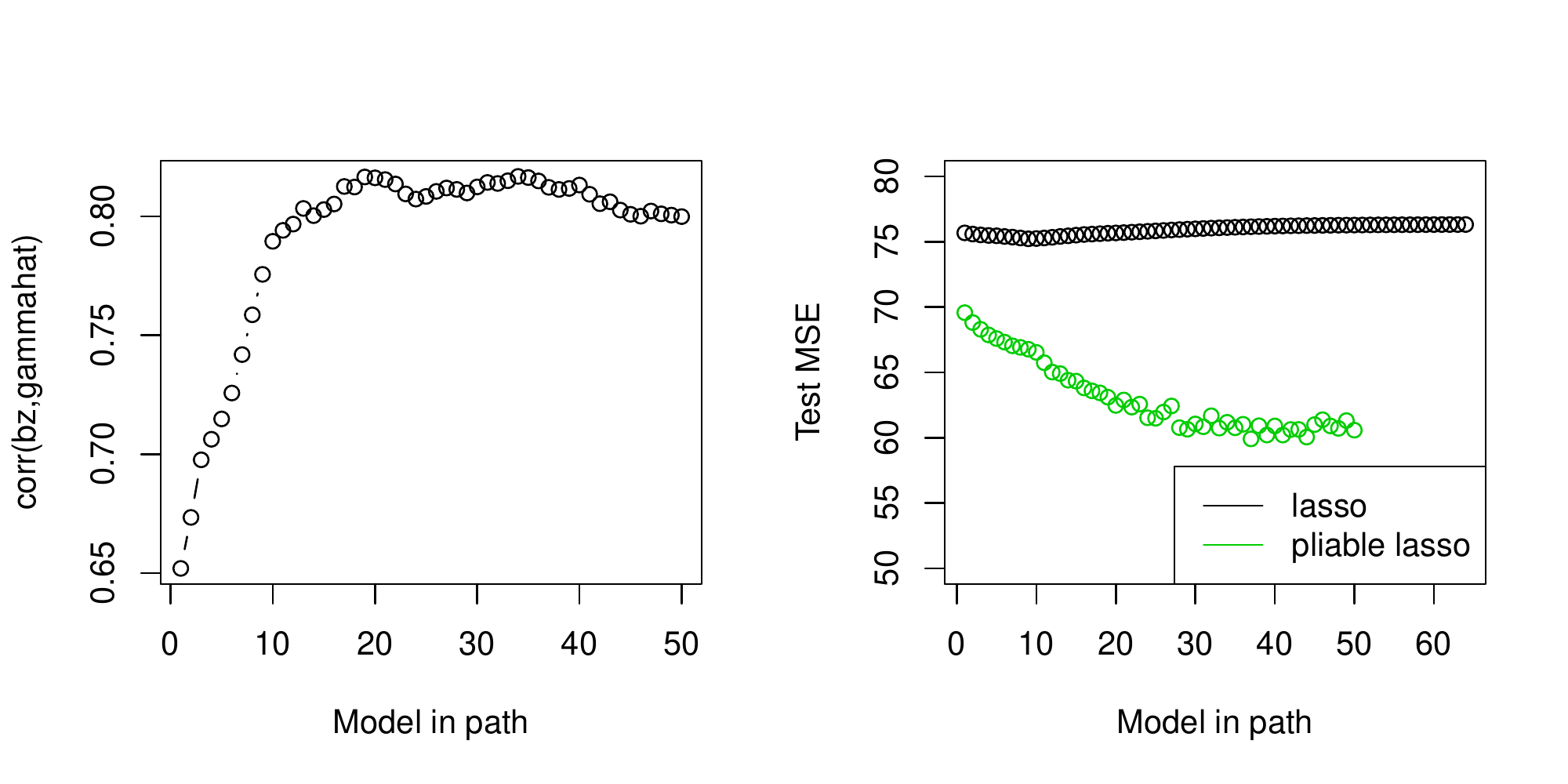}
\caption[fig:unkZ]{\em Results for pliable lasso, with modifying variables $Z$ unobserved. The left panel shows the correlation between the estimated weights $\hat\gamma$ and the generating
weights $b_z$. The right panel  shows the test error for the pliable lasso (green) compared to the usual lasso (black).
}
\label{fig:unkZ}
\end{figure}

Data were generated as $Pr(z=0|x) =1/(1+\exp(-x^T b_z))$ with $b=(0,0,\ldots 0,10,10,-10,10)$.
The Bayes error for classifying $z$ from $x$ was 35\%.  We set $\sigma=0.25$ giving an SNR for $y$  of about 1.4.
Figure \ref{fig:unkZ} shows the result of applying just two cycles of the above procedure, starting with $\gamma$ equal to the least squares
estimate of $y$ on $X$. The left panel shows the correlation between the estimated weights $\hat\gamma$ and the generating
weights $b_z$. The right panel  shows the test error for the pliable lasso (green) compared to the usual lasso (black).
We see that in this idealized example, there is potential for learning the modifying variables from the data itself.

\section{Further topics}
\label{sec:further}
\begin{description}
\item{(a)} {\em Tree-based pliable lasso.}  A  more flexible version of the pliable lasso can be derived by  using a regression tree to estimate the $\Delta_j(Z)$ factors in  model \ref{eqn:plasso}.
This allows for more general  and higher order interactions between  $X$ and $Z$.
A coordinate descent algorithm can be derived for the resulting optimization, using a weighted regression tree fit.
We have experimented with this idea with some success, but the non-convexity of the objective function makes it difficult
to work with.
\item{(b}) {\em Extensions to other models.} The ideas presented here for the Gaussian regression model
can be extended to other settings such as generalized linear models and the Cox proportional hazards  model. One  use can the standard Newton-style  approach employed by the {\tt glmnet} program, solving a weighted problem in the inner loop.  

In the Cox model, $Z$ could be a set of modifying variables, as in the Gaussian model of this paper. But one could also use a categorical-valued $Z$ to denote the strata in a stratified Cox model.
In more detail, the stratified  Cox model assumes that the hazard function in the $k$th stratum has the form
\begin{equation}
h_k(t|x)= h_{0k}(t) e^{x^T \beta}
\end{equation}
where $h_{0k}(t)$ is the baseline hazard function for the $k$th stratum. The log-partial likelihood is typically used for
estimation, and is a sum over strata. We can generalize this model to
\begin{equation}
h_k(t|x)= h_{0k}(t) \exp[\sum_{j=1}^p x_j (\beta_j +\theta_{jk} I(Z=k))].
\end{equation}
This would allow the effects of some features to vary over strata. In a similar way, one could choose $Z$ to index the risk sets in a survival analysis or the matched sets in a conditional logistic regression.

\item {(c)} {\em Screening rules}. A number of authors have proposed  variable screening rules for speeding up the
coordinate descent algorithm for the lasso and related procedures.  These include  \cite{laurent}, \cite{TBFHSTT2012},
\cite{wang12:_lasso_screen_rules_via_dual_polyt_projec} and \cite{Ndiaye2016}.  Since the objective function for the
pliable lasso is closely related to the lasso and sparse group lasso, we are optimistic that effective screening rules
could be derived for its optimization.
\end{description}

\medskip

An R package for the pliable lasso will be made available in the CRAN library

\medskip

{\bf Acknowledgements:} We'd like to thank Jacob Bien and Leying Guan for helpful comments.  Robert Tibshirani was supported by
NIH grant 5R01 EB001988-16 and NSF grant 19 DMS1208164.

 \section*{Appendix: details of the optimization}

The pliable lasso model has the form

$$\hat y= \beta_0\one +Z\theta_0+ \sum_{j=1}^p (X_j\beta_j+ W_j\theta_j)$$
where $W_j=X_j\circ  Z$ (elementwise multiplication in each column). 

We note that  $\beta_0$ and $\theta_0$  are optional components in the model, and are unpenalized when included. Hence for simplicity of notation we omit them from consideration:
at the beginning of each pass over the predictors,  we estimate $\beta_0$ and $\theta_0$ by regressing the current residual on  $(1,Z$).
In that case  the outcome $y$ appearing below can be replaced by   $y-\hat\beta_0-Z\hat\theta_0$.

The objective function is
$$J(\beta, \theta)=\frac{1}{2N} \sum( y_i-\hat y_i)^2   + (1-\alpha) \lambda \sum_{j=1}^p\Bigl((||(\beta_j,\theta_j)||_2 +
||\theta_j||_2 \Bigr)  +\alpha \lambda\sum_{j,k} |\theta_{jk}|_1.$$
The subgradient equations are 
{N}
\begin{eqnarray}
\frac{dJ}{d\beta_j} &=&-\frac{1}{N} X_j^Tr + (1-\alpha)\lambda u=0 \cr
\frac{dJ}{d\theta_j} &=&-\frac{1}{N} W_j^Tr +  (1-\alpha)\lambda (u_2+ u_3) + \alpha\lambda v=0
\end{eqnarray}
where $r=y-\hat y$,   and

$u= \frac{\beta_j}{||\beta_j,\theta_j||_2}$ if $(\beta_j, \theta_j)\neq 0$ and $\in \{ u: ||u||_2 \leq 1\}$  if  $(\beta_j,\theta_j)= 0$

$u_2= \frac{\theta_j}{||(\beta_j,\theta_j)||_2}$ if $(\beta_j, \theta_j) \neq 0$ and $\in \{u_2: ||u_2||_2 \leq 1\}$  if  $(\beta_j,\theta_j)= 0$

$u_3= \frac{\theta_j}{||\theta_j ||_2}$ if  $\theta_j \neq 0$ and $\in \{u_3: ||u_3||_2 \leq 1\}$  if  $\theta_j= 0$

$v\in {\rm sign}(\theta_j)$

\bigskip

{\em Screening conditions: } Define the partial residual, leaving out the $j$th group, as
$$r_{(-j)}=y-\sum_{\ell \neq j}  ([X_\ell\hat\beta_\ell+W_\ell\hat\theta_\ell] $$

Then $(\hat\beta_j,\hat\theta_j)=0$  if
\begin{equation}
|X_j^Tr_{(-j)}/N| \leq (1-\alpha)\lambda\; {\rm  and} \;
||S(W_j^Tr_{(-j)}/N,\alpha\lambda)||_2 \leq 2(1-\alpha)\lambda.
\label{cond1}
\end{equation}

Otherwise we check if $\hat\beta_j\neq 0, \hat\theta_j=0$ by first computing
\begin{equation}
\hat\beta_j= (n/||X_j||_2^2)\cdot S(X_j^Tr_{-(j)}/N,(1-\alpha)\lambda)
\label{bhat}
\end{equation}

and then checking if
\begin{equation}
||S(W_j^T(r_{(-j)}- X_j\hat\beta_j)/N,\alpha\lambda)||_2 \leq (1-\alpha)\lambda.
\label{cond2}
\end{equation}

{\em Iterations:}  (if $(\hat\beta_j,\hat\theta_j) \neq 0$):

\medskip

 Let $\gamma_j=(\beta_j,\theta_j)$. The majorized objective function is

 $$M(\gamma_j)=\ell_0+(\gamma_j-\gamma_0)^T\nabla \ell + \frac{1}{2t} ||\gamma_j-\gamma_0||_2^2 +(1-\alpha) \lambda( || \gamma_j ||_2 +||\theta_j||_2)+ \alpha \lambda\sum_{j,k} |\theta_{jk}|_1$$
 with $\ell_0=\nabla\ell (r_{(-j)},\gamma_0)= (-X_j^Tr_{(-j)}/N, -W_j^Tr_{(-j)}/N$ for squared error loss.

This is equivalent to minimizing
$$\tilde M(\gamma_j)=\frac{1}{2t}||\gamma_j-\gamma_0-t\nabla \ell )||_2^2 +(1-\alpha) \lambda (|| \gamma_j ||_2 +||\theta_j||_2)+ \alpha \lambda\sum_{j,k} |\theta_{jk}|_1.$$

Then $\hat\beta_j,\hat\gamma$ satisfy
\begin{eqnarray}
\Bigl( 1+\frac{t(1-\alpha)\lambda}{||\hat\gamma_j||_2}\Bigr) \hat\beta_j&=&\beta_0-t\nabla_{\beta_j}\ell \cr
\Bigl( 1+t(1-\alpha)\lambda (\frac{1}{||\hat\theta_j||_2}+\frac{1}{||\hat\gamma_j||_2})\Bigr) \hat\theta_j &=&S(\theta_0-t\nabla_{\theta_j}\ell, t\alpha\lambda).
\end{eqnarray}

Let $a=||\hat\beta_j||_2, b=||\hat\theta_j||_2$ so that $||\hat\gamma_j||_2=\sqrt{a^2+b^2}$. Take the norm of both sides in each equation above giving
\begin{eqnarray}
\Bigl( 1+\frac{t(1-\alpha)\lambda}{ \sqrt{a^2+b^2}}\Bigr) a&=&|\beta_0-t\nabla_{\beta_j}\ell |\cr
\Bigl( 1+t(1-\alpha)\lambda(\frac{1}{b}+\frac{1}{\sqrt{a^2+b^2}) })\Bigr) b &=&||S(\theta_0-t\nabla_{\theta_j}\ell, t\alpha\lambda)||_2.
\label{normeqn}
\end{eqnarray}
Defining $c=t(1-\alpha)\lambda$, $g_1=|\beta_0-t\nabla_{\beta_j}\ell |, g_2=||S(\theta_0-t\nabla_{\theta_j}\ell, t\alpha\lambda)||_2$, let $r_1, r_2$ be the roots of the quadratic equation $r^2+2c_1r-2c_1g_2-g_1^2-g_2^2$. Then
\begin{eqnarray}
\hat a =\frac{g_1 u'}{c+u''} ;\; \hat b =\frac{g_1 v'(c-g_2)}{c+v'}
\label{normsoln}
\end{eqnarray}
where $u',u'',v',v''$ are each equal to one of the roots $r_1,r_2$ to satisfy (\ref{normeqn}).

Finally, the solutions $\hat\beta_j, \hat\theta_j$ satisfy
\begin{eqnarray}
\Bigl( 1+\frac{t(1-\alpha)\lambda}{\sqrt{\hat a^2+\hat b^2}}\Bigr) \hat\beta_j&=&\beta_0-t\nabla_{\beta_j}\ell \cr
\Bigl( 1+t(1-\alpha)\lambda(\frac{1}{\hat b}+\frac{1}{\sqrt{\hat a^2+\hat b^2}) })\Bigr) \hat\theta_j &=&S(\theta_0-t\nabla_{\theta_j}\ell, t\alpha\lambda)
\end{eqnarray}
Letting $c_1,c_2$ be the constants multiplying $\hat\beta_j$ and $\hat\theta_j$ above, we have the update equations
\begin{eqnarray}
\hat\beta_j&=&\frac{\beta_0-t\nabla_{\beta_j}\ell}{c_1}\cr
\hat\theta_j&=&\frac{\||S(\theta_0-t\nabla_{\theta_j}\ell, t\alpha\lambda)||_2}{c_2}.
\label{updateeqn}
\end{eqnarray}
We use this to define updates $(\hat\beta_j,\hat\theta_j) \leftarrow U(\hat\beta_j^{old},\hat\theta_j^{old}, t)$

All of this leads to the procedure given in Algorithm 1 below.

\bigskip
\setcounter{algorithm}{0}
\begin{algorithm}
\caption{ (Detailed) Algorithm for the Pliable Lasso}

\begin{description}
\item For a decreasing path of $\lambda$  values:
\begin{description}
\item Repeat until convergence:
\begin{enumerate}
\item Compute $\hat\beta_0$ and $\hat\theta_0$  from  the  least squares regression of the current residual on $Z$.
\item For predictor  $k=1,2,\ldots p,1,2,\ldots $ 
\begin{description}
\item{(a)}  Check condition (\ref{cond1})  for $(\hat\beta_j, \hat\theta_j)=0$. if zero, skip to next $k$
\item{(b)} Otherwise,  compute $\hat\beta_j$ from (\ref{bhat}) and then check if $\hat\theta_j=0$ from (\ref{cond2}). If zero, skip to next $k$
\item{(c)} Otherwise, if both $\hat\beta_j, \hat\theta_j$ are nonzero:
 \item Repeat until convergence
\begin{description}
   \item{(d)} Solve equation (\ref{normeqn}) for the norms $a,b$  using  (\ref{normsoln})
   \item{(e)} Update  $(\hat\beta_j,\hat\theta_j) \leftarrow U(\hat\beta_j^{old},\hat\theta_j^{old}, t)$ from (\ref{updateeqn}).
   (Nesterov acceleration and backtracking can be added  for speed and to ensure convergence).
  \end{description}
  \end{description}
\end{enumerate}
\end{description}
\end{description}
\end{algorithm}

\newpage
\bibliographystyle{agsm}
\bibliography{tibs.bib}
 \end{document}